\documentstyle[aps,pra,preprint]{revtex}
\begin{document}
\draft

\title{Semiclassical theory of transport in antidot lattices\thanks{This paper
is dedicated to Prof.\
H.\ Wagner on the occasion of his 60th birthday.}}
\author{Gregor Hackenbroich and Felix von Oppen}
\address{Max-Planck-Institut f\"ur Kernphysik, 69117 Heidelberg, Germany}
\date{\today}
\maketitle

\begin{abstract}
Motivated by a recent experiment by Weiss et al.\ [Phys.\ Rev.\ Lett.\ {\bf
70}, 4118 (1993)],
we present a detailed study of quantum transport in large antidot arrays whose
classical
dynamics is chaotic. We calculate the longitudinal and Hall conductivities
semiclassically starting from the Kubo formula. The leading contribution
reproduces the
classical conductivity. In addition, we find oscillatory quantum corrections to
the classical
conductivity which are given in terms of the periodic orbits of the system.
These periodic-orbit
contributions provide a consistent explanation of the quantum oscillations in
the
magnetoconductivity observed by Weiss et al. We find that the phase of the
oscillations
with Fermi energy and magnetic field is given by
the classical action of the periodic orbit. The amplitude is determined by the
stability
and the velocity correlations of the orbit. The amplitude also decreases
exponentially with
temperature on the scale of the inverse orbit traversal time $\hbar/T_\gamma$.
The Zeeman
splitting leads to beating of the amplitude with magnetic field. We also
present an analogous
semiclassical derivation of Shubnikov-de Haas oscillations where the
corresponding classical
motion is integrable. We show that the quantum oscillations in antidot lattices
and the
Shubnikov-de Haas oscillations are closely related. Observation of both effects
requires
that the elastic and inelastic scattering lengths be larger than the lengths of
the relevant
periodic orbits. The amplitude of the quantum oscillations in antidot lattices
is of a higher power in Planck's constant $\hbar$ and hence smaller than that
of
Shubnikov-de Haas oscillations. In this sense, the quantum oscillations in the
conductivity
are a sensitive probe of chaos.
\end{abstract}

\narrowtext
\section{Introduction}

Over the past few years high-mobility semiconductor heterostructures have
proved to be a valuable tool for studying the implications of quantum chaos
experimentally \cite{chaos}. Advances in technology have made it possible to
manufacture
structures whose elastic and inelastic mean free paths exceed the relevant
sample dimensions.
In these {\it ballistic} (micro)structures essentially all large-angle
scattering is due to the
geometric features of the device \cite{Webb}. Hence, they can be used in
particular for experimental
studies of billiard systems \cite{Weiss1,Marcus,Weiss2,Keller,Chang} which have
been
central to the theoretical development of quantum chaos \cite{chaos}. These
novel
experimental realizations of quantum-chaotic systems are important because
there are
actually only a small number of systems in the field of quantum chaos that
allow for a
controlled comparison between theory and experiment.

Much of the work on quantum chaos has focused on the spectral properties
of systems whose classical dynamics is chaotic \cite{chaos}. However, the
detailed
spectrum is not an accessible observable in ballistic semiconductor structures.
Rather, one measures thermodynamic properties such as the magnetic
susceptibility \cite{Levy} and the persistent current \cite{Mailly} or
transport
properties such as the conductance \cite{Weiss1,Marcus,Weiss2,Keller,Chang}
of the sample. Hence the challenge
to theory was whether these quantities are sensitive to the
nature of the underlying classical dynamics. This question has now been
answered in the affirmative for both thermodynamic and transport properties.
With regard to transport properties
of ballistic microstructures, it has been predicted that the
power spectrum of the conductance fluctuations \cite{Jalabert,Baranger2} and
the
lineshape of the weak-localization peak \cite{Baranger2,Baranger1} are probes
of
quantum chaos and there is accumulating experimental evidence in favor of these
predictions \cite{Marcus,Keller,Chang}. There are also
predictions that the amplitudes of the persistent current \cite{Oppen1} and the
magnetization \cite{Oppen2,Ullmo} are sensitive probes of quantum chaos. These
have
not yet been subject to experimental tests.

Chaotic transport in semiconductor structures has been investigated in two
kinds
of experiments. Experiments on {\it mesoscopic} ballistic
cavities with the shape of chaotic billiards coupled to leads
\cite{Marcus,Keller,Chang} have focused
on generic effects such as conductance fluctuations and weak localization.
These effects are familiar from the study of transport in diffusive conductors
\cite{Lee,Altshuler,Weaklocalization}.
In the other type of experiment one measures magnetotransport in {\it
macroscopic} antidot
lattices. Antidot lattices consist of a periodic array of holes ``drilled''
into a
two-dimensional electron gas \cite{Weiss1,Weiss2,Nakamura}.
They are {\it macroscopic} in the
sense that the phase-coherence length is much smaller than the system size
(the lattice spacing of the antidot array, on the other hand, may be smaller
than
the phase-coherence length) \cite{Schuster}.
Antidot lattices can be viewed as an approximate experimental realization of
the
Sinai billiard \cite{Sinai} which is a well-known chaotic system.
Experiments on antidot arrays were mostly concerned with features in the
magnetotransport which reflect the specific nature of the sample. In this paper
we present a theoretical study of transport in antidot lattices \cite{Hacken}.

Two analytical methods dominated theoretical studies of quantum chaos,
random-matrix theory \cite{Bohigas} and the semiclassical approximation
\cite{Gutzwiller}.
While the first is most appropriate for describing generic effects due to long
trajectories,
semiclassical theory is especially useful in describing effects due to short
system-specific
trajectories. We find that semiclassical theory is particularly suited to
address the
system-specific features in the magnetoconductivity of antidot lattices.
In the context of quantum chaos interest in semiclassical theory derives mostly
from the
fact that, strictly speaking, chaos is a concept of classical physics.
Semiclassical
theory allows one to study most explicitly the implications of (classical)
chaos in
quantum mechanics. In the context of ballistic structures semiclassical
theory is a natural approach because the Fermi wavelength is often the shortest
length scale
in the problem. Furthermore, the contributions
of long classical trajectories are frequently suppressed by finite temperature
or inelastic
scattering, rendering the semiclassical expressions quite manageable
\cite{Oppen1,Oppen2,Ullmo}.

The first experiments on antidot lattices by Weiss et al.\ \cite{Weiss1}
focused on the
regime of classical transport. They revealed a series of pronounced
magnetoresistance peaks
at magnetic fields where the classical cyclotron radius is commensurate with
the lattice
spacing of the antidot array. For these magnetic-field values, the electrons
can circle
around a certain number of antidots for long times. This effectively slows them
down
leading to an increase in the resistivity \cite{Weiss1,Fleischmann,runaway}.
Here we are concerned with
more recent experiments at lower temperatures \cite{Weiss2} which revealed
additional quantum
oscillations in the magnetoresistance superimposed on the classical peaks. The
period
of these oscillations corresponds to roughly a flux quantum threading a unit
cell of the
antidot lattice. Depending on the sample these oscillations varied between
$1/B$ periodicity
familiar from Shubnikov-de Haas oscillations \cite{Ando} and $B$  periodicity
familiar from
Aharonov-Bohm oscillations \cite{Washburn}.  Weiss et al.\ \cite{Weiss2}
explained
these  observations phenomenologically by assuming that the resistivity is
proportional to  the square of the density of states, $\rho_{xx}\sim
d^2(E_F,B)$.
The period of the oscillations can then be understood in terms of a
few short periodic orbits which, according to Gutzwiller's trace formula
\cite{Gutzwiller}, give oscillatory contributions
to the density of states. This approach is clearly unsatisfactory because it
does not
make any predictions concerning the absolute amplitude of the oscillations and,
more
importantly, because this relation between the resistivity and the density of
states
cannot be derived from first principles. In fact, this relation certainly does
not
reproduce the pronounced classical peaks in the resistivity \cite{bdep}.

Nevertheless, the apparent success of the phenomenological theory of Ref.\
\cite{Weiss2}
suggested that there is indeed a relation between quantum corrections to
transport properties
and classical periodic orbits. In this paper we present a semiclassical
theory for the conductivity starting from first principles \cite{Silberbauer}.
Since we are considering
macroscopic
antidot lattices much larger than elastic and inelastic mean free paths we
start from
the Kubo formula. To leading order in Planck's constant $\hbar$ we reproduce
the classical
Kubo formula. According to Fleischmann et al.\ \cite{Fleischmann}
this contribution accounts for the classical peaks
in the resistivity. The central result of this paper is that we find quantum
corrections to the classical conductivity which explain the additional quantum
oscillations observed by Weiss et al.\ \cite{Weiss2}.
These quantum contributions to the conductivity are indeed given in terms of
classical periodic
orbits. The contribution of each orbit oscillates as function of Fermi energy
and
magnetic field with a phase equal to its classical action. The amplitude
depends on
the stability of and velocity correlations along the orbit. The predictions of
our
theory for the period, the amplitude, and the temperature dependence of the
periodic-orbit contributions are consistent with the experimental results of
Weiss
et al.\ \cite{Weiss2}. It was previously known that periodic orbits are
important in describing the oscillatory corrections to the density of states
\cite{Gutzwiller} and hence for thermodynamic properties of mesoscopic systems
\cite{Argaman1,Oppen1,Oppen2,Ullmo,Agam,Prado}. Our results combined with the
experimental
results of Ref.\ \cite{Weiss2} show that periodic  orbits are also important
for
quantum corrections of transport properties \cite{Wilkinson}. There have also
been
attempts to interpret the ``frequency spectrum'' of conductance fluctuations in
billiard-shaped
ballistic microstructures in terms of periodic orbits \cite{Marcus,Schuster}.
The relevance of our
results to these experiments is not clear because the conductance of such
phase-coherent samples
is more appropriately described using the Landauer-B\"uttiker approach rather
than the
Kubo formula.

We find that the quantum oscillations in the conductivity of ``chaotic''
antidot lattices are
closely related to the Shubnikov-de Haas oscillations in the conductivity of
the ``classically
integrable''
two-dimensional electron gas. To make this relation explicit we present an
analogous
semiclassical derivation of Shubnikov-de Haas oscillations. Even though both
effects are
closely related, the technical treatment is somewhat different.
It is well known that integrable and chaotic systems must be treated
differently in
the semiclassical theory of the spectral density \cite{Gutzwiller}.
Likewise we find that certain spatial integrals which must be computed exactly
in the
derivation of Shubnikov-de Haas oscillations are evaluated by the
stationary-phase method
for chaotic antidot lattices. As a result, the amplitude of Shubnikov-de Haas
oscillations
is larger than that of the quantum oscillatons in antidot lattices. In analogy
to the
semiclassical theory of the spectral density \cite{Gutzwiller}, our results
for antidot lattices are strictly valid only if the classical dynamics is
completely
chaotic. In section \ref{sec:comp} we present evidence that this is indeed the
case to
a good approximation for the main sample investigated in Ref.\ \cite{Weiss2}.

The outline of the paper is as follows. In section \ref{sec:sdh} we present the
semiclassical
theory for Shubnikov-de Haas oscillations. Section \ref{sec:anti} contains the
central results
of this paper on the longitudinal conductivity (\ref{sec:long}) and the Hall
conductivity
(\ref{sec:hall}) of antidot lattices,
and a comparison between our theory and the experimental results of Ref.\
\cite{Weiss2} (\ref{sec:comp}). Various details of the calculations in sections
\ref{sec:sdh},
\ref{sec:long}, and \ref{sec:hall} are given in three appendices. Finally, we
conclude in
section \ref{sec:con}.

\section{Shubnikov-de Haas oscillations}
\label{sec:sdh}
\narrowtext
In this section we present a semiclassical derivation of the Shubnikov-de Haas
oscillations
in the magnetoconductivity of two-dimensional electron gases. Our motivation
for doing so is
twofold. First, it is instructive to see how the semiclassical methods
which we employ in our derivation of the  quantum oscillations in antidot
lattices
work in a more familiar context.  Second, it turns out that the Shubnikov-de
Haas
and the quantum oscillations  in antidot lattices are closely related
phenomena. This
relationship is brought out explicitly by comparing the calculations for both
effects.

Throughout this paper we consider transport in {\it macroscopic} systems in
the  sense that the sample dimensions are much larger than the elastic and
inelastic
mean free paths. In this regime, transport is appropriately described by the
Kubo
formula. Here, we start from the Kubo formula for the longitudinal conductivity
\begin{equation}
  \sigma_{xx}={e^2\pi\hbar\over\Omega}\,{\rm Tr}\left\{{\hat v}_x\delta_
\Gamma(E_F- {\hat H}){\hat v}_x\delta_\Gamma(E_F-{\hat H})\right\},
\label{Kubo}
\end{equation}
where ${\hat H}=(1/2m)({\hat{\bf p}}
-e{\bf A({\hat{\bf r}}}))^2$ is the Hamiltonian of a two-dimensional electron
gas in
a perpendicular magnetic field, ${\bf A}=(-By/2,Bx/2)$ is the vector potential
taken in the
symmetric gauge, and ${\hat v}_x$ is the $x$  component of the
velocity operator (here $m$ denotes the {\it effective} mass of the electrons).
The Fermi energy is denoted by $E_F$ and the  volume of the
system by $\Omega$. It is well known that one needs to include some disorder
scattering
when directly computing the dc conductivity within the Kubo formalism
\cite{disorder}.
We include weak disorder at the level of the Born approximation
by giving the $\delta$-functions a finite width $\Gamma=\hbar/2\tau_{el}$
($\tau_{el}$
is the elastic scattering time). This is sufficient to prove our main point,
i.e., that
there is a precise connection between classical periodic orbits and quantum
corrections
to the conductivity. However, it should be kept in mind that the Born
approximation cannot
fully describe smooth disorder potentials because it does not distinguish
between the
elastic and the transport mean free paths. Unfortunately, it appears to be a
difficult
problem to go beyond the Born approximation when using the semiclassical
approach
presented in this paper. In principle, the level broadening $\Gamma$ also
includes
a contribution due to inelastic scattering. This will be left implicit
throughout
the paper.

To evaluate the Kubo formula (\ref{Kubo}), we start by expressing the
$\delta$-functions in Eq.\ (\ref{Kubo}) in terms of the retarded and advanced
Green  functions ${\hat G}^{\pm}(E)=[E-{\hat H}\pm i\Gamma]^{-1}$,
\begin{equation}
  \delta_\Gamma(E_F-{\hat H})=-{1\over2\pi i}[{\hat G}^+(E_F)-{\hat G}^-(E_F)].
\label{deltagreen}
\end{equation}
Evaluating the trace in (\ref{Kubo}) in position representation, one finds that
the
longitudinal conductivity can be expressed as
\begin{equation}
  \sigma_{xx}=\sigma_{xx}^{++}+\sigma_{xx}^{--}+\sigma_{xx}^{+-},
\end{equation}
where
\begin{equation}
   \sigma_{xx}^{\pm\pm}=-{e^2\pi\hbar\over\Omega}\!\left({1\over2\pi
m}\right)^2\!
    \int\!d^2rd^2r^\prime\!\left[\!\left({\hbar\over
    i}\nabla_x-eA_x({\bf r})\right)\!G_{{\bf r},{\bf
    r}^\prime}^\pm (E_F)\right]\!\!\left[\!\left({\hbar\over
    i}\nabla_x^\prime-eA_x({\bf r}^\prime)\right)\!G_{{\bf
    r}^\prime,{\bf r}}^\pm (E_F)\right]
\label{adad}
\end{equation}
and
\begin{equation}
   \sigma_{xx}^{+-}=2{e^2\pi\hbar\over\Omega}\!\left({1\over2\pi m}\right)^2\!
    \int\!d^2rd^2r^\prime\!\left[\!\left({\hbar\over
    i}\nabla_x-eA_x({\bf r})\right)\!G_{{\bf r},{\bf
    r}^\prime}^-(E_F)\right]\!\!\left[\left({\hbar\over
    i}\nabla_x^\prime-eA_x({\bf r}^\prime)\right)\!G_{{\bf
    r}^\prime,{\bf r}}^+(E_F)\right]\!.
\label{Kubogreen}
\end{equation}
\narrowtext
\noindent Here, $G^\pm_{{\bf r},{\bf r}^\prime}(E)=\langle{\bf r}|{\hat
G}^\pm(E)|{\bf r}^\prime
\rangle$ denotes the matrix elements of the Green functions. It is shown in
appendix \ref{app:++}
that both $\sigma_{xx}^{++}$ and $\sigma_{xx}^{--}$ vanish in the semiclassical
approximation. Hence, we focus on $\sigma_{xx}^{+-}$ in the remainder of this
paper.

Having expressed the longitudinal conductivity in terms of Green functions, we
now
approximate these by their semiclassical expressions. Generally, the
semiclassical Green function
$G^+_{{\bf r}^\prime,{\bf r}}(E)$ is a sum over all classical trajectories of
energy
$E$ from ${\bf r}$ to ${\bf r}^\prime$. For free electrons in a magnetic
field, there are an infinite number of trajectories connecting ${\bf r}$ to
${\bf
r}^\prime$ if the  distance between ${\bf r}$ and ${\bf r}^\prime$ is less than
twice
the cyclotron radius $R_c$: Two different cyclotron orbits labeled by $S$ (for
short)
and $L$ (for long) which can each be  traversed completely for $n=0,1,2\ldots$
times
pass through both points (cf.\ Fig.\ 1).  There are no trajectories from
${\bf r}$ to ${\bf r}^\prime$ when $|{\bf r}-{\bf r}^\prime|>2R_c$ and the
semiclassical Green function vanishes for such arguments. An explicit
calculation
shown in appendix \ref{app:eg} gives \widetext
\begin{equation}
  G_{{\bf r}^\prime,{\bf r}}^+(E)={m\over2i\hbar}\sum_{n=0}^\infty
  \sum_{q=S,L}\left({\omega_c\over\pi i\hbar E|\sin(\omega_c
T_{n,q})|}\right)^{1/2}
\exp\left(-{T_{n,q}\over2\tau_{el}}\right)\exp\left\{{i\over\hbar}
   S_{n,q}-{i\pi\over2}
  \eta_{n,q}\right\}.
\label{GreenSdH}
\end{equation}
Here $T_{n,q}$ denotes the traversal time of the trajectory $(n,q)$,
\begin{equation}
   T_{n,q}={2\pi n\over\omega_c}+t_q,
\label{Tnq}
\end{equation}
where
\begin{eqnarray}
  t_S &=&{2\over\omega_c}\arcsin(|{\bf r}-{\bf r}^\prime|/2 R_c)
  \nonumber\\
  t_L &=&{2\over\omega_c}[\pi-\arcsin(|{\bf r}-{\bf r}^\prime|/2R_c)]
\label{times}
\end{eqnarray}
are the times for going from ${\bf r}$ to ${\bf r}^\prime$ without completely
traversing
the cyclotron orbit ($\omega_c=eB/m$ is the cyclotron frequency). Each
trajectory contributes
with a phase factor given by its classical action
\widetext
\begin{equation}
  S_{n,q}(E)=ET_{n,q}+{m\omega_c\over2}\left[ {|{\bf r}-{\bf r}^\prime|^2 \over
  2}\cot\left({\omega_cT_{n,q}\over2}\right)-(xy^\prime-x^\prime y)\right]\!
\label{action}
\end{equation}
\narrowtext
and the Maslov indices $\eta_{n,S} =2n$, $\eta_{n,L}=2n+1$. The corresponding
expression for
$G^-_{{\bf r},{\bf r}^\prime}(E)$ follows from the relation
\begin{equation}
  G^-_{{\bf r},{\bf r}^\prime}(E)=[G^+_{{\bf
  r}^\prime,{\bf r}}(E)]^*.
\end{equation}
One notes that both Green functions appearing in (\ref{Kubogreen}) are given in
terms of
trajectories from ${\bf r}$ to ${\bf r}^\prime$.

Substituting these semiclassical expressions for the Green functions into Eq.\
(\ref{Kubogreen})
for the conductivity one has a double sum over classical trajectories $(n,q)$
from ${\bf r}$ to
${\bf r}^\prime$,
\widetext
\begin{eqnarray}
  \sigma_{xx}&=&{\omega_c\over2\Omega E_F}\left({me\over2\pi\hbar}\right)^2
  \sum_{n_1=0}^{\infty}\sum_{n_2=0}^{\infty}
  \sum_{q_1=S,L} \sum_{q_2=S,L}\int d^2r\int d^2r^\prime\,
  (v_{q_1})_x(v^\prime_{q_2})_x
  \nonumber \\
  & &\times{\exp\{-(T_{n_1,q_1}+T_{n_2,q_2})/2\tau_{el}\}\over
    |\sin(\omega_c T_{n_1,q_1})\sin(\omega_c T_{n_2,q_2})|^{1/2}}
  \exp\left\{{i\over\hbar}[S_{n_2,q_2}-S_{n_1,q_1}]-{\pi\over2}[\eta_{n_2,q_2}-
  \eta_{n_1,q_1}]\right\},
\label{sigdouble}
\end{eqnarray}
\narrowtext
\noindent
where we have used the relations
\widetext
\begin{eqnarray}
  {1\over m}\left({\hbar\over i}\nabla_x-eA_x({\bf r})\right)S_{n,q}
  &=&-(v_q)_x,
  \\
  {1\over m}\left({\hbar\over i}\nabla_x^\prime-eA_x({\bf r}^\prime)
  \right) S_{n,q}&=&(v_q^\prime)_x.
\label{velocities}
\end{eqnarray}
\narrowtext
\noindent
Here $(v_q)_x$ and $(v_q^\prime)_x$ denote the $x$ components of the velocities
at the
points ${\bf r}$ and ${\bf r}^\prime$, respectively. One notes that for
$q_1=q_2$, the
difference in action $S_{n_2,q_2}-S_{n_1,q_1}$ is independent  of both ${\bf
r}$ and
${\bf r}^\prime$. Only these terms contribute to leading order in the
semiclassical limit
because the integrand oscillates rapidly when $q_1\ne q_2$. A further
simplification
arises because both the times $T_{n,q}$ and the difference in  action depend
only on
$|{\bf r}-{\bf r}^\prime|$. It is useful to introduce center-of-mass
coordinates and
to change variables to the times $t_q$, \widetext \begin{eqnarray}
  \int_\Omega d^2r\int_{|{\bf r}-{\bf r}^\prime|\leq2R_c}d^2r^
  \prime&=&2\pi\Omega\int_0^{2R_c}|{\bf r}-{\bf r}^\prime|\,
  d|{\bf r}-{\bf r}^\prime|
  \nonumber \\
  &=&2\pi\Omega\omega_c R_c^2\left[\int_0^{\pi/\omega_c}dt_S\,|\sin(
  \omega_c t_S)|+\int_{\pi/\omega_c}^{2 \pi/\omega_c} dt_L\,|\sin(
  \omega_c t_L)|\right].
\label{volelement}
\end{eqnarray}
\narrowtext
\noindent
Using $(v_q)_x(v^\prime_q)_x=v_x(0)v_x(t_q)$ and combining the integrations
over $t_S$ and $t_L$,
one obtains
\widetext
\begin{eqnarray}
  \sigma_{xx}&=&{e^2m\over2\pi\hbar^2}
  \int_0^{2\pi/\omega_c}dt\,v_x(0)v_x(t)\,e^{-t/2\tau_{el}}
  \nonumber \\
  & & \times\sum_{n_1,n_2=0}^{\infty}\exp\left\{-{\pi\over\omega_c\tau_{el}}
(n_1+n_2)\right\}\cos\left\{2\pi\left({E_F\over\hbar\omega_c}-
{1\over2}\right)(n_2-n_1)
  \right\}.
\label{sigSdH1}
\end{eqnarray}
\narrowtext
Finally performing the sum over $n_1+n_2$ and introducing $p=|n_1-n_2|$ we find
\begin{equation}
  \sigma_{xx}={2e^2\over
h}\,\left({E_F\over2\pi\hbar\omega_c}\right)\,C(v_x,v_x)
  \left[1+2\sum_{p=1}^{\infty}
\exp\left(-{p\pi\over\omega_c\tau_{el}}\right)\cos\left\{2\pi p
  \left({E_F\over\hbar\omega_c} -{1\over2}\right)\right\}\right].
\label{sigSdH2}
\end{equation}
Here we introduced the correlation function
\begin{equation}
  C(v_x,v_x)={1\over R_c^2}\int_0^{2\pi/\omega_c}dt\int_0^\infty
dt^\prime\,v_x(t)
  v_x(t+t^\prime)\,e^{-t^\prime/\tau_{el}}.
\label{correlation}
\end{equation}
This way of writing the result will be useful for comparison with the
conductivity of
the antidot lattice derived in the next section. For comparison with the
standard result
for Shubnikov-de Haas oscillations in the literature \cite{Ando} one easily
computes
\widetext
\begin{equation}
  C(v_x,v_x)={\pi\omega_c\tau_{el}\over1+(\omega_c\tau_{el})^2}.
\label{corrSdH}
\end{equation}
\narrowtext
Using that the density of electrons (per spin) is $n_e=mE_F/2\pi\hbar^2$ the
prefactor in
Eq.\ (\ref{sigSdH2}) can be rewritten in the more familiar form $(n_e
e^2\tau_{el}/m)
[1+(\omega_c\tau_{el})^2]^{-1}$. This proves that the semiclassical
approximation
reproduces the low-field approximation ($E_F\gg\hbar\omega_c$)
to the quantum-mechanical result (within
the Born approximation) \cite{Ando}. We note that the classical Drude
conductivity
arises from the diagonal terms in the double sum over trajectories in Eq.\
(\ref{sigdouble}). The Shubnikov-de Haas oscillations arise from non-diagonal
terms
corresponding to different numbers $n_1$ and $n_2$ of complete revolutions
around
the same cyclotron orbit. In the next section we will find a closely analogous
structure of the terms contributing to the conductivity of antidot lattices.

\section{Quantum transport in antidot lattices}
\label{sec:anti}

\subsection{Longitudinal conductivity}
\label{sec:long}

In the following we consider the longitudinal conductivity of macroscopic
antidot
lattices.  We start from the Kubo formula (\ref{Kubo}) with the Hamiltonian
${\hat H}=(1/2m)({\hat{\bf p}}-e{\bf A})^2+V({\hat{\bf r}})$ where $V({\hat{\bf
r}})$
is the antidot potential. A semiclassical approximation to the Green function
$G^+_{{\bf r}^\prime,{\bf r}}(E)$ for general potentials $V({\bf r})$ is
obtained by
making consistent use of stationary-phase integration both for the path
integral for the propagator and for the Fourier transform to the energy
domain. One finds that the Green-function matrix elements are approximated
by sums over the classical trajectories $\gamma({\bf r},{\bf r}^\prime)$ of
energy $E$ going from ${\bf r}$ to ${\bf r}^\prime$ \cite{Gutzwiller},
\begin{equation}
    G^+_{{\bf r}^\prime,{\bf r}}(E)={1\over(2\pi)^{1/2}}\left({1\over
      i\hbar}\right)^{3/2}\sum_{\gamma({\bf r},{\bf r}^\prime)}
      D_\gamma({\bf r},{\bf r}^\prime)
      \exp\left\{{i\over\hbar}S_\gamma({\bf r},
      {\bf r}^\prime;E)-i{\pi\over2}\eta_\gamma\right\}.
\label{Green}
\end{equation}
The phase of the Green function is determined by the classical action
$S_\gamma({\bf r},{\bf r}^\prime;E)=\int_\gamma d{\bf r}\,{\bf p}$ of
the trajectory $\gamma$ (${\bf p}$ is the canonical momentum and the integral
is taken along the trajectory) and by the Maslov index $\eta_\gamma$, which
counts the number of conjugate points along the trajectory. Each trajectory
contributes with an amplitude $D_\gamma$ which is determined by disorder and
the
classical phase-space density,
\begin{equation}
   D_\gamma({\bf r},{\bf r}^\prime)=\exp\left(-{T_\gamma\over2\tau_{el}}\right)
      \left|\begin{array}{cc}
      {\textstyle\partial^2 S_\gamma\over\textstyle\partial r^\prime_i\partial
      r_j} &
      {\textstyle\partial^2 S_\gamma\over\textstyle\partial E\partial r_j} \\
      {\textstyle\partial^2 S_\gamma\over\textstyle\partial r^\prime_i\partial
      E} &
      {\textstyle\partial^2 S_\gamma\over\textstyle\partial^2E} \end{array}
      \right|^{1/2}.
\label{greenamp}
\end{equation}
Here the traversal time of the trajectory is denoted by $T_\gamma$.
The corresponding expression for $G^-_{{\bf r},{\bf r}^\prime}(E)$
follows from $G^-_{{\bf r},{\bf r}^\prime}(E)=[G^+_{{\bf r}^\prime,{\bf r}}
(E)]^*$. Importantly, both Green functions appearing in (\ref{Kubogreen})
are given in terms of trajectories running from ${\bf r}$ to ${\bf r}^\prime$.
The main difference between the semiclassical expressions for $G^+_{{\bf
r}^\prime,
{\bf r}}(E)$ and $G^-_{{\bf r},{\bf r}^\prime}(E)$ is that the actions
enter in the exponent with opposite signs.

Inserting the semiclassical expressions for the Green functions into Eq.\
(\ref{Kubogreen}) one finds that $\sigma_{xx}$ is given by a double sum
over classical trajectories from ${\bf r}$ to ${\bf r}^\prime$. In the
simplest approximation one retains only the diagonal terms in this double
sum for which the rapidly-oscillating phase factors drop out,
\begin{equation}
  \sigma_{xx}\simeq2{e^2\pi\hbar\over\Omega}\left({1\over2\pi\hbar}\right)^3
  \int d^2rd^2r^\prime\sum_{\gamma({\bf r},{\bf r}^\prime)} v_x v_x^\prime
  D^2_\gamma({\bf r},{\bf r}^\prime).
\label{diagonal}
\end{equation}
Here we used the elementary relations
\begin{eqnarray}
        {1\over m}\left({\hbar\over i}\nabla_x-eA_x({\bf r})\right) S_{\gamma}
        ({\bf r},{\bf r}^\prime;E_F)&=&-v_x
\label{vx}
        \\
        {1\over m}\left({\hbar\over i}\nabla_x^\prime-eA_x({\bf
r}^\prime)\right)
        S_{\gamma}({\bf r},{\bf r}^\prime;E_F)&=&v_x^\prime,
\label{vxprime}
\end{eqnarray}
where $v_x$ and $v_x^\prime$ denote the $x$ components of the initial and
final velocities, respectively. This contribution gives the classical
conductivity. To see this, we multiply the right-hand side of Eq.\
(\ref{diagonal}) by $1=\int dE\,\delta(E-E_F)$ and introduce as new
integration variables the initial position ${\bf r}$, the initial momentum
${\bf p}$, and the traversal time $t$ of the trajectory $\gamma$. The
corresponding Jacobian is given by
\begin{equation}
  d^2r\,d^2p\,dt=d^2r\,d^2r^\prime\,dE\sum_{\gamma({\bf r},{\bf r}^\prime)}
  \left|\begin{array}{cc}
      {\textstyle\partial p_i\over\textstyle\partial r^\prime_j}&
      {\textstyle\partial p_i\over\textstyle\partial E} \\
      {\textstyle\partial t\over\textstyle\partial r^\prime_j}&
      {\textstyle\partial t\over\textstyle\partial E} \end{array}
      \right|.
\label{jacobian}
\end{equation}
The sum over $\gamma$ appears because initial conditions (${\bf r}$,${\bf p}$)
specify a classical trajectory uniquely while boundary conditions (${\bf r}$,
${\bf r}^\prime$) can be satisfied by more than one trajectory. One notes that
the Jacobian (\ref{jacobian}) involves precisely the same determinant which
also
appears in the amplitude factors $D_\gamma({\bf r},{\bf r}^\prime)$ in the
Green functions (\ref{Green}). Hence, one finds for the conductivity
\begin{equation}
   \sigma_{xx}\simeq {e^2\over h^2\Omega}\int d^2r\,d^2p\,dt\, v_x(0)v_x(t)
   \,e^{-t/\tau_{el}} \, \delta(E_F-H({\bf p},{\bf r})) \, ,
\end{equation}
where $H({\bf p},{\bf r})$ denotes the classical Hamiltonian. Introducing the
classical
phase-space averages
\begin{equation}
   \langle f({\bf p},{\bf r})\rangle_\Gamma={1\over h^2\Omega N(0)} \int
d^2r\,d^2p
   \,f({\bf p},{\bf r})\,\delta(E_F-H({\bf p},{\bf r})),
\end{equation}
($N(0)=m/2\pi\hbar^2$ is the density of states at the Fermi energy in two
dimensions)
the conductivity can be rewritten in the familiar form
\begin{equation}
  \sigma_{xx}\simeq e^2N(0)\int_0^\infty dt\,\langle
v_x(0)v_x(t)\rangle_\Gamma\, e^{-t/\tau_{el}}.
\label{classkubo}
\end{equation}
This is precisely the {\it classical} Kubo formula. Hence, the diagonal
approximation to the double sum over classical trajectories in Eq.\
(\ref{Kubogreen})
is equivalent to the classical approximation. At first sight, it may appear
inconsistent that the classical limit (\ref{classkubo}) still -- via the
density
of states $N(0)$ -- involves Planck's constant $\hbar$. However, this is simply
a
consequence of the fact that we use the Fermi-Dirac distribution instead of the
Boltzmann distribution appropriate for a complete classical limit.

As shown by Fleischmann et al.\ \cite{Fleischmann} the classical Kubo formula
(\ref{classkubo}) explains the peaks in the magnetoresistance when the
cyclotron radius and the lattice spacing of the antidots are commensurate.
In this paper, we are mainly interested in the quantum corrections
$\delta\sigma_{xx}$ to the classical conductivity. Clearly, they must arise
from
the nondiagonal terms in the double sum in (\ref{Kubogreen}),
\begin{eqnarray}
  \delta\sigma_{xx}={e^2\over h^2\Omega}& &\,\int
d^2r\,d^2r^\prime\,\sum_{\gamma_1
  ({\bf r},{\bf r}^\prime)\ne\gamma_2({\bf r},{\bf r}^\prime)}(v_1)_x
(v_2^\prime)_x
  D_{\gamma_1}({\bf r},{\bf r}^\prime)D_{\gamma_2}({\bf r},{\bf r}^\prime)
  \nonumber\\ & &\times
  \exp\left\{{i\over\hbar}\left[S_{\gamma_2}({\bf r},{\bf r}^\prime;E_F)
  -S_{\gamma_1}({\bf r},{\bf
r}^\prime;E_F)\right]-{i\pi\over2}\left[\eta_{\gamma_2}
  -\eta_{\gamma_1}\right]\right\}.
\end{eqnarray}
In these terms, the phase factors do not cancel and hence the integrand
involves
rapidly-oscillating factors. Within the semiclassical approach for completely
chaotic
systems, it is natural to evaluate the spatial integrals by the
stationary-phase
method. This procedure becomes exact in the semiclassical limit $\hbar\to0$.
Considering the contribution arising from trajectories $\gamma_1({\bf r},{\bf
r}^\prime)$
and $\gamma_2({\bf r},{\bf r}^\prime)$, the stationary-phase conditions for the
${\bf r}$ and ${\bf r}^\prime$ integrations are
\begin{eqnarray}
        {\bf \nabla}[S_{\gamma_2}({\bf r},{\bf r}^\prime;E_F)
        -S_{\gamma_1}({\bf r},{\bf r}^\prime;E_F)]&=&{\bf p}_1-{\bf p}_2=0
\label{saddle1}
        \\
        {\bf \nabla}^\prime[S_{\gamma_2}({\bf r},{\bf r}^\prime;E_F)
        -S_{\gamma_1}({\bf r},{\bf r}^\prime;E_F)]&=&{\bf p}_2^\prime-{\bf p}
        _1^\prime =0.
\label{saddle2}
\end{eqnarray}
Here, ${\bf p}_i$ and ${\bf p}_i^\prime$ denote the initial and final
(canonical) momenta of trajectory $\gamma_i$. Since the vector
potential cancels from the stationary-phase conditions, one finds that the
integrand becomes stationary when initial and final velocities of $\gamma_1$
and $\gamma_2$ are identical. Since also the initial and final positions of
the two trajectories must each coincide one might at first think that this
implies
$\gamma_1=\gamma_2$ which would just be the classical contribution
discussed above. However, there is an additional possibility to
satisfy the stationary-phase conditions (\ref{saddle1},\ref{saddle2})
when the two trajectories are part of a classical periodic orbit.
Then $\gamma_1$ and $\gamma_2$ can differ by an integer number $p$ of
traversals of the orbit as illustrated in Fig.\ 2. In fact, for each
(primitive) periodic orbit $\gamma$ there is an infinite set of pairs of
trajectories
$\gamma_1$ and $\gamma_2$ which satisfy the stationary-phase condition
(\ref{saddle1},
\ref{saddle2}). The pairs are labeled by $p$ which was defined above
and by $n$ which counts the number of times the shorter trajectory from ${\bf
r}$
to ${\bf r^\prime}$ returns to its starting point ${\bf r}$. It is these
quantum corrections to the conductivity which we focus on in the following.

The evaluation of the stationary-phase integrations over ${\bf r}$ and
${\bf r}^\prime$, while different in detail, is closely analogous to performing
the final trace in the derivation of Gutzwiller's trace formula
\cite{Gutzwiller}.
First we note that the saddle-point action is just given by the action
$S_\gamma(E_F)$
of the (primitive) periodic orbit,
\begin{equation}
   S_{\gamma_2}({\bf r},{\bf r}^\prime;E_F)-S_{\gamma_1}({\bf r},{\bf
r}^\prime;E_F)
                 =pS_\gamma(E_F).
\end{equation}
To evaluate the integral over the quadratic fluctuations around the
stationary-phase
point it is useful to
introduce a new set of orthogonal coordinates ${\bf r}=(z,y)$ where $z$
measures
the arclength along the periodic orbit $\gamma$ and $y$ measures the transverse
deviations from the orbit (cf.\ Ref.\ \cite{Gutzwiller}).
The Jacobian of this transformation is unity,
$d^2r\,d^2r^\prime=dz\,dy\,dz^\prime
\,dy^\prime$. While the integrals over $y$ and $y^\prime$ can be done
by the stationary-phase method, the integrals along the periodic orbit must be
performed exactly because the second variation of the action vanishes in this
direction. Introducing the matrix $F$ of second derivatives of
$S_{21}=S_{\gamma_2}-
S_{\gamma_1}$ with respect to $y$ and $y^\prime$,
\begin{equation}
  F=\left[\begin{array}{cc}
      {\textstyle\partial^2_y S_{21}}&
      {\textstyle\partial_y\partial_{y^\prime} S_{21}} \\
      {\textstyle\partial_{y^\prime}\partial_y S_{21}}&
      {\textstyle\partial^2_{y^\prime}S_{21}} \end{array}
      \right]
\end{equation}
(here we used the shorthand notation $\partial_y=\partial/\partial y$) and the
vector
${\bf y}^T=[y,y^\prime]$ of transverse deviations from the periodic orbit, one
finds
for the fluctuation integral
\begin{equation}
   \int dy\,dy^\prime\,\exp\left\{{i\over 2\hbar}{\bf y}^TF{\bf y}\right\}
   ={2\pi i\hbar\over\sqrt{\det F}}.
\end{equation}
A straight-forward but tedious calculation presented in appendix
\ref{app:stability}
shows that this fluctuation amplitude combined with the semiclassical
amplitudes
of the Green functions can be expressed in terms of the monodromy matrix
$M_\gamma$
of the periodic orbit alone,
\begin{equation}
   {D_{\gamma_1}({\bf r},{\bf r}^\prime)D_{\gamma_2}({\bf r},{\bf
r}^\prime)\over
   \sqrt{|\det F|}}={\exp\{-(T_{\gamma_1}+T_{\gamma_2})/2\tau_{el}\}\over |{\bf
v}(z)|\,
   |{\bf v}(z^\prime)|}\,|{\rm Tr}M^p_\gamma-2|^{-1/2}.
\label{monostab}
\end{equation}
Finally, we simplify the integrations along the periodic orbit over $z$ and
$z^\prime$. Since the integrations over ${\bf r}$ and ${\bf r}^\prime$ are over
the volume of the system, the integrations over $z$ and $z^\prime$ extend
over a full period around the {\it primitive} periodic orbit. Furthermore,
since ${\bf r}$
is the starting and ${\bf r}^\prime$ the end point of the trajectories, the
limits of integration
are $0\le z\le L_\gamma$ and $z\le z^\prime \le z^\prime+L_\gamma$, where
$L_\gamma$
is the length of the (primitive) periodic orbit. Hence, the $z,z^\prime$
integrations
become
\begin{eqnarray}
\sum_{n=0}^\infty\int_0^{L_\gamma}&dz&\int_z^{z+L_\gamma}
dz^\prime{v_x(z)v_x(z^\prime)\over
  |{\bf v}(z)|\,|{\bf
v}(z^\prime)|}\exp\{-(T_{\gamma_1}+T_{\gamma_2})/2\tau_{el}\}
  \nonumber\\
  & &=\exp(-pT_\gamma/2\tau_{el})
  \sum_{n=0}^\infty\int_0^{T_\gamma}dt\int_0^{T_\gamma}dt^\prime
v_x(t)v_x(t+t^\prime)
  \exp\{-(nT_\gamma+t^\prime)/\tau_{el}\}
  \nonumber\\
  & &=\exp(-pT_\gamma/2\tau_{el})\int_0^{T_\gamma}dt\int_0^\infty dt^\prime
  v_x(t)v_x(t+t^\prime)\exp(-t^\prime/\tau_{el}).
\label{zzprimeint}
\end{eqnarray}
Here, $T_\gamma=L_\gamma/v_F$ denotes the period of the primitive periodic
orbit.
Due to the translational symmetry of the antidot lattice each periodic orbit
appears
$\Omega/a^2$ times ($a$ is the lattice spacing). Including only a single
representative
of each class of periodic orbits in the sum over $\gamma$, we find the final
result
\begin{equation}
   \delta\sigma_{xx}={2e^2\over h}{R_c^2\over a^2}
   \sum_\gamma\sum_{p=1}^\infty\exp(-pT_\gamma/2\tau_{el})
   C_\gamma(v_x,v_x)\,{\cos\{pS_\gamma(E_F)/\hbar-\pi p\alpha_\gamma/2\}\over
   |{\rm Tr}M^p_\gamma-2|^{1/2}},
\label{statphas}
\end{equation}
where the sum over $\gamma$ is over all primitive periodic orbits and $p$
counts the
repeated traversals. We also introduced the correlation function
\begin{equation}
   C_\gamma(v_x,v_x)={1\over R_c^2}\int_0^{T_\gamma}dt\int_0^\infty dt^\prime
v_x(t)
   v_x(t+t^\prime)\exp(-t^\prime/\tau_{el}).
\label{velcor}
\end{equation}
Some periodic orbits of antidot lattices turn out to be quite close to
circular. For these
it can be a useful approximation to replace the present correlation function by
its
analog (\ref{correlation}) for Shubnikov-de Haas oscillations.
By collecting all phases one convinces oneself that the Maslov index
$\alpha_{\gamma}$
appearing in Eq.\ (\ref{statphas}) is the same that enters Gutzwiller's trace
formula
\cite{Gutzwiller}. Hence, periodic-orbit contributions to the conductivity and
to the
density of states have the same phase which is physically reasonable. The
result
(\ref{statphas}) for the periodic-orbit corrections to the conductivity of
antidot
lattices should be compared to the result (\ref{sigSdH2}) for Shubnikov-de Haas
oscillations.
The principal difference is that
the stability of the periodic orbit expressed in terms of the monodromy matrix
enters in
the case of chaotic dynamics. Furthermore, one notes that the quantum
oscillations for the
antidot lattices are of a higher power in Planck's constant $\hbar$ and hence
smaller in the
semiclassical limit than the Shubnikov-de Haas oscillations.

At finite temperatures the Kubo formula takes on the form
\begin{equation}
  \sigma_{xx}=-{e^2\pi\hbar\over\Omega}\int_0^\infty dE\,{\partial
f_\mu(E)\over
  \partial E}{\rm Tr}\left\{{\hat v}_x\delta_
  \Gamma(E- {\hat H}){\hat v}_x\delta_\Gamma(E-{\hat H})\right\},
\label{Kubot}
\end{equation}
where $\mu$ denotes the chemical potential and $f_\mu(E)$ is the Fermi-Dirac
distribution
$f_\mu(E)=[1+\exp(\beta(E-\mu))]^{-1}$ ($\beta=1/k_BT$ is the inverse
temperature). To
compute the temperature dependence of the periodic-orbit contribution to the
conductivity
one notes that the dominant energy dependence (to leading order in $\hbar$) is
in the
phase factor. Hence, we need to compute the integral
\begin{equation}
  I(T)=-\int_0^\infty dE\,{\partial f_\mu(E)\over\partial
E}\,\cos\left\{{p\over\hbar}S_\gamma(E)
  -{\pi \over2}p\alpha_\gamma\right\}.
\end{equation}
Since the derivative of the Fermi function is strongly peaked at the Fermi
energy,
we extend the lower limit of integration to $-\infty$ and expand the action to
first
order around the Fermi energy,
\begin{equation}
  I(T)={\rm
Re}{\partial\over\partial\mu}\exp\left\{{i\over\hbar}pS_\gamma(\mu)-{i\pi\over
2}
  p\alpha_\gamma\right\}\int_{-\infty}^\infty
dE\,f_\mu(E)\,\exp\left\{{i\over\hbar}pT_\gamma
  (E-\mu)\right\}.
\end{equation}
Here we used that $dS_\gamma/dE=T_\gamma$.
This integral can be evaluated by contour integration. The Fermi-Dirac
distribution has
poles at $E=\mu+i(2n+1)\pi/\beta$ with residues $-1/\beta$. One finds
\begin{equation}
  I(T)=\cos\left\{{p\over\hbar}S_\gamma(\mu)-{\pi\over2}p\alpha_\gamma\right\}
  {(\pi pT_\gamma/\hbar\beta)\over\sinh(\pi pT_\gamma/\hbar\beta)}.
\end{equation}
Thus we find for the periodic-orbit contribution to the conductivity at finite
temperature
\begin{eqnarray}
   \delta\sigma_{xx}={2e^2\over h}{R_c^2\over a^2}
   \sum_\gamma\sum_{p=1}^\infty& &\exp(-pT_\gamma/2\tau_{el})C_\gamma(v_x,v_x)
   \nonumber\\
   & &\times{(\pi pT_\gamma/\hbar\beta)\over\sinh(\pi pT_\gamma/\hbar\beta)}\,
   {\cos\{pS_\gamma(E_F)/\hbar-\pi p\alpha_\gamma/2\}\over
   |{\rm Tr}M^p_\gamma-2|^{1/2}}.
\label{statphasT}
\end{eqnarray}
The contribution from each periodic orbit decreases exponentially with
temperature on
the scale of $\hbar/T_\gamma$. In principle, there is an addional temperature
dependence
due to the variation of the inelastic scattering length with temperature.
However, this
temperature dependence is slow compared to that arising from thermal smearing.
For this reason,
Eq.\ (\ref{statphasT}) does indeed describe the dominant temperature dependence
of the effect.

The Zeeman coupling of the electronic spin to the magnetic field is readily
incorporated
into our results. It effectively leads to separate Fermi energies
$E_F\pm\mu_BB$ for the
two spin components (here $\mu_B=e\hbar/2m_e$ is the Bohr magneton with $m_e$
the bare
electron mass). Expanding the action to
linear order in the Zeeman energy one finds that it leads to beating of the
oscillation
amplitude,
\begin{eqnarray}
   \delta\sigma_{xx}={2e^2\over h}& &{R_c^2\over a^2}
   \sum_\gamma\sum_{p=1}^\infty\exp(-pT_\gamma/2\tau_{el})C_\gamma(v_x,v_x)
   \nonumber\\
   & &\times\cos(pT_\gamma\mu_BB/\hbar)\,
   {(\pi pT_\gamma/\hbar\beta)\over\sinh(\pi pT_\gamma/\hbar\beta)}\,
   {\cos\{pS_\gamma(E_F)/\hbar-\pi p\alpha_\gamma/2\}\over
   |{\rm Tr}M^p_\gamma-2|^{1/2}}.
\label{statphasbeat}
\end{eqnarray}
The corresponding magnetic-field scale is $\hbar/\mu_BT_\gamma$.

\subsection{Hall conductivity}
\label{sec:hall}

The calculation for the Hall conductivity $\sigma_{xy}$ is closely analogous
to that for the longitudinal conductivity described in the previous section.
The Hall conductivity can be expressed in terms of two contributions
\cite{Streda}
\begin{equation}
  \sigma_{xy}=\sigma_{xy}^I+\sigma_{xy}^{II},
\label{sighall}
\end{equation}
where
\begin{equation}
  \sigma_{xy}^I={ie^2\hbar\over2\Omega}{\rm Tr}\left\{{\hat v_x}{\hat G}^+(E_F)
  {\hat v_y}\delta_\Gamma(E_F-H)-{\hat v_x}\delta_\Gamma(E_F-H){\hat v_y}{\hat
G}^-(E_F)\right\}
\end{equation}
and
\begin{equation}
  \sigma_{xy}^{II}=-{e\over\Omega}\int_0^{E_F}dE\,{\partial\over\partial
B}\,\rho(E).
\end{equation}
Here, $\rho(E)$ denotes the density of states. First, we evaluate the
contribution
$\sigma_{xy}^{II}$ using Gutzwiller's trace formula which gives
a semiclassical approximation to the density of states $\rho(E)$. The Weyl
contribution to the density of states is independent of the magnetic field.
Thus,
we only need
the oscillatory contributions $\delta\rho(E)$ to the density of states which
can be
expressed in terms of a sum over the periodic orbits $\gamma$ of the system,
\begin{equation}
   \delta\rho(E)={\Omega\over
a^2}{1\over\pi\hbar}\sum_{\gamma}\sum_{p=1}^\infty
   {T_\gamma\exp(-pT_\gamma/2\tau_{el})\over|{\rm
Tr}M_\gamma^p-2|^{1/2}}\cos\left\{
   {1\over\hbar}pS_\gamma(E)-{\pi\over2}p\alpha_\gamma\right\}.
\end{equation}
Noting that to leading order in Planck's constant $\hbar$ we only need to keep
the
magnetic-field and energy dependence in the oscillatory factor, one readily
finds
\begin{equation}
   \delta\sigma_{xy}^{II}=-{2e^2\over
h}\sum_{\gamma}\sum_{p=1}^\infty\exp(-pT_\gamma/2\tau_{el})
   \left({1\over a^2}{\partial S_\gamma(E_F)\over\partial
B}\right){\cos\{pS_\gamma(E_F)/
   \hbar-\pi p\alpha_\gamma/2\}\over |{\rm Tr}M_\gamma^p-2|^{1/2}}.
\label{hallII}
\end{equation}
We now proceed to evaluate the first contribution $\sigma_{xy}^I$. The
evaluation of
this term is completely analogous to that for $\sigma_{xx}^{+-}$. We do not
consider the
classical Hall conductivity and focus on the periodic-orbit contribution right
away.
Again, we only need to consider contributions from products of advanced and
retarded
Green functions.
By analogy with the evaluation of $\sigma_{xy}^{+-}$ we can immediately write
down the
result of the calculation for the Hall conductivity,
\begin{equation}
    \delta\sigma_{xy}^I={2e^2\over h}{R_c^2\over a^2}
    \sum_\gamma\sum_{p=1}^\infty
         \exp(-pT_\gamma/2\tau_{el})\, C_\gamma(v_x,v_y)\,
         {\cos\{pS_\gamma(E_F)/\hbar-p\pi\alpha_\gamma/2\}\over|{\rm Tr}
         M_\gamma^p-2|^{1/2}},
\label{hallI}
\end{equation}
where the velocity correlations are given by
\begin{equation}
   C_\gamma(v_x,v_y)={1\over R_c^2}\int_0^{T_\gamma}dt\int_0^\infty
      dt^\prime\,v_x(t)v_y(t+t^\prime)\,
       e^{-t^\prime/\tau_{el}}.
\label{velcorxy}
\end{equation}
\noindent The Maslov index $\alpha_\gamma$ is again the same that also appears
in Gutzwiller's
trace formula.
Finite temperature and the Zeeman splitting lead to the same additional factors
which also appeared in the treatment of the longitudinal conductivity in the
preceding
section. For an approximately circular periodic orbit of radius $R$ traversed
at constant angular
velocity $\omega$ one finds
\begin{equation}
    \delta\sigma_{xy}^{\rm cir}\approx{2e^2\over h}\,{\pi R^2\over a^2}\,
    {1\over1+(\omega\tau_{el})^2}\,\exp(-pT_\gamma/2\tau_{el})\,
     {\cos[pS_\gamma(E_F)/\hbar-p\pi\alpha_\gamma/2]\over|{\rm Tr}
     M_\gamma^p-2|^{1/2}}.
\end{equation}
The same disorder-dependent prefactor appears in the theory of Shubnikov-de
Haas oscillations
\cite{Ando}. Note that the amplitude of a circular orbit in the Hall
conductivity is smaller
by a factor $1/\omega\tau_{el}$ than that in the longitudinal conductivity.

\subsection{Comparison with experiment}
\label{sec:comp}

Weiss et al.\ \cite{Weiss2,Weiss3} measured both the longitudinal conductivity
and the
Hall conductivity of macroscopic antidot lattices as function of
magnetic field. The antidot lattices were patterned from high-mobility
GaAs-AlGaAs
heterostructures with elastic and inelastic mean free paths of order $5\mu$m.
For the main sample discussed below the lattice spacing was $a=200$nm and the
antidot diameter $d=100$nm. The Fermi wavelength was of order $\lambda_F=50$nm.

At higher temperatures (typically $T=4.7$K) Weiss et al.\ \cite{Weiss1} found a
number
of peaks in the longitudinal resistivity at magnetic fields where the classical
cyclotron
radius is commensurate with the lattice spacing of the antidot array. This
phenomenon has been discussed theoretically by Fleischmann et al.\
\cite{Fleischmann}.
Here we are concerned with the additional quantum oscillations in both
longitudinal
and Hall conductivities superimposed on the classical conductivity. These were
observed by
Weiss et al.\ \cite{Weiss2} at subkelvin temperatures. The experimental results
for
the main sample investigated in Ref.\ \cite{Weiss2} are shown in Fig.\ 3.
To focus on the quantum oscillations, we have plotted
$\delta\sigma=\sigma(T=0.4{\rm K})
-\sigma(T=4.7{\rm K})$. In the following we compare our theoretical results to
the experimental
data for this particular sample. The quantum oscillations have a period of the
order of
one flux quantum per unit cell and the oscillations in $\sigma_{xx}$ and
$\sigma_{xy}$ are
in phase. One reads off from Fig.\ 3 that the amplitude is roughly
$\delta\sigma_{xx}\approx\delta\sigma_{xy}\approx0.015({\rm k}\Omega)^{-1}$.

A detailed comparison between theory and experiment is complicated by a number
of
factors. Corrections to the semiclassical results may be relevant because the
Fermi
wavelength is not much smaller than the antidot radius. In addition, our
approximations
are strictly valid only for completely chaotic systems. By contrast, the
experimental system
exhibits mixed dynamics in part of the magnetic-field range over which the
quantum
oscillations have been observed. One expects that this does {\it not} affect
the
phase of the oscillations strongly but does have an influence on their
amplitude. A detailed
comparison of theory and experiment also requires use of a model for the
antidot
potential which is not known a priori. Perhaps the most serious shortcoming is
our
treatment of disorder which is sufficient to prove that there is a rigorous
connection
between quantum corrections to the conductivity and classical periodic orbits
but which is not fully adequate to describe the smooth disorder potentials
typical of semiconductor heterostructures. For want of a better treatment
we have considered the elastic mean free path as a fit parameter, keeping it
merely within reasonable bounds.

We model the antidot potential by \cite{Fleischmann}
\begin{equation}
    V({\bf r})=V_0\left[\cos(\pi x/a)\cos(\pi y/a)\right]^{\beta}.
\label{potential}
\end{equation}
Here, $V_0$ controls the diameter of the antidots at the Fermi energy and
$\beta$ fixes
the steepness of the antidot potential. Hence, all parameters but $\beta$ are
fixed
directly by experiment.
Following Ref.\ \cite{Weiss2} we choose $\beta=2$. In Ref.\ \cite{Weiss2}
this choice was motivated by the phase of the quantum oscillations. Using the
approach
of Fleischmann et al.\ \cite{Fleischmann} we have also computed numerically the
classical
resistivity for this potential. Experimentally one finds a single peak located
at $R_c=a/2$.
As shown in Fig.\ 4, the model potential with $\beta=2$ also leads
to a single dominant peak at the same magnetic-field value in the classical
resistivity.
This lends additional support to this choice of $\beta$. Other choices such as
$\beta=4$
lead to considerable additional structure in the resistivity
\cite{Fleischmann}.

To study the nature of the classical dynamics for this choice of model
potential we
have generated Poincare surfaces of section. For $\beta=2$ and magnetic fields
such
that $R_c=a/2$ one finds that phase space is entirely chaotic (within our
resolution).
This is shown in Fig.\ 5(a). The classical dynamics becomes more (less)
chaotic as one decreases (increases) the magnetic field from $R_c=a/2$. In
particular,
islands of regular motion appear for larger magnetic fields.
Our results for completely chaotic systems should be applicable for magnetic
fields
with $R_c>a/2$. However, as already mentioned above, one might expect a
somewhat reduced
level of accuracy because the trajectories are close to stability.
The classical dynamics also becomes less chaotic as one increases the
steepness of the antidot potential. This is shown in Fig.\ 5(b) for
$\beta=4$ and $R_c=a/2$, where one finds a large regular island in the Poincare
section.

Our central result Eq.\ (\ref{statphas}) expresses the quantum oscillations in
the
conductivity in terms of the classical periodic orbits of the antidot lattice.
Since
long orbits are suppressed due to elastic and inelastic scattering, stability
amplitude,
and finite temperature, we focus on the contributions from the shortest
periodic orbits
labeled by (a), (b), and (c) in Fig.\ 6. It turns out that the action of orbit
(a) changes only slowly with magnetic field giving rise to oscillations with
large period.
These are not resolved experimentally because of the small magnetic-field
range.
Note that while orbit (b) has fourfold symmetry, orbit (c) has only twofold
symmetry
so that the latter contributes twice. As already pointed out by Weiss et al.\
\cite{Weiss2}
these orbits explain the phase of the quantum oscillations. Both orbits have
nearly the same
action and one finds that the predicted minima from Eq.\ (\ref{statphas})
indicated by arrows
in Fig.\ 3 agree well with the observed minima in the conductivity.
Furthermore,
our results predict in agreement with experiment that longitudinal and Hall
conductivity
are in phase.

The amplitude of each periodic orbit $\gamma$ in Eq.\ (\ref{statphas}) depends
on the traversal
time $T_\gamma$, the velocity correlations $C_\gamma(v_x,v_x)$, and the
stability factor
$|{\rm Tr}M_\gamma-2|^{-1/2}$. Our numerical results for these quantities at
the classical
peak $R_c=a/2$ and for $\omega_c\tau_{el}=2$ are collected in table I. Orbit
(c) in fact
consists of three periodic orbits in close vicinity of each other. In view of
the Fermi
wavelength $\lambda_F$ we do not believe that it is justified to include all
three orbits
separately. Instead we only include one of the three with typical stability.
Presumably
this tends to underestimate the amplitude and hence a more accurate treatment
would lead to
closer agreement between theory and experiment: Using the numbers of table I
we find $\delta\sigma_{xx}\approx0.008({\rm k}\Omega)^{-1}$ from Eqs.\
(\ref{statphas})
and $\delta\sigma_{xy}\approx0.004({\rm k}\Omega)^{-1}$ from Eq.\ (\ref{hallI})
\cite{mean}.
These theoretical estimates are smaller by factors of two and four,
respectively, than the
experimental results. We believe that this order-of-magnitude agreement is
reasonable
given the theoretical uncertainties mentioned at the beginning of this section.

Our results predict a beat period due to the Zeeman splitting which is rather
large
compared to the magnetic-field range over which the quantum oscillations were
observed.
While the effects of the Zeeman splitting were thus unobservable, the predicted
temperature
dependence is consistent with experimental results. The amplitude of the
quantum
oscillations is expected to decrease exponentially with temperature on the
scale $\hbar/\pi
T_\gamma$. Hence, one predicts a characteristic temperature of 0.7K. This is
consistent
with experiment where quantum oscillations were observed at 0.4K but not at
4.7K
\cite{Weiss1,Weiss2}.

In the introduction we remarked that the periodicity of the quantum
oscillations varies
between $1/B$ periodicity and $B$ periodicity depending on the experimental
sample
\cite{Weiss2}. To understand this observation, consider the magnetic-field
range
where the cyclotron radius is of the order of half the lattice spacing so that
electrons can circle around a single antidot. If the antidot diameter is small
the relevant periodic orbits are only weakly perturbed by the presence of the
antidot potential. Hence, one expects $1/B$ periodicity familiar from the
Shubnikov-de
Haas oscillations in unpatterned samples. On the other hand, if the antidot
diameter
is larger, the periodic orbits are modified by the antidot potential. In
particular,
they will be inhibited in contracting with increasing magnetic field. Hence,
with increasing antidot diameter one crosses
over to a situation where the enclosed area of the periodic orbits is
approximately
constant with magnetic field. In this limit the change of the classical action
with
magnetic field is merely due to the Aharonov-Bohm flux and hence one finds $B$
periodicity.

\section{Discussion and summary}
\label{sec:con}

Motivated by a recent experiment by Weiss et al.\ \cite{Weiss2} we have studied
the quantum
transport in large antidot lattices. We have evaluated the Kubo formula for the
longitudinal
and the Hall conductivity
semiclassically for systems whose classical dynamics is chaotic. While the
leading contribution
of the semiclassical expansion is just the classical Kubo formula as expected,
we have identified
quantum corrections due to the classical periodic orbits of the system. The
contribution from
each periodic orbit oscillates as function of the Fermi energy and the magnetic
field with the phase
given by the classical action of the orbit. The amplitude is determined by the
stability and
the velocity correlations of the orbit. Furthermore, we find that the
periodic-orbit contributions
to the conductivity decrease with increasing temperature on the scale
$\hbar/T_\gamma$
where $T_\gamma$ denotes the traversal time of the orbit. Our results provide a
consistent
explanation for the quantum oscillations superimposed on the classical
magnetoresistivity
of antidot lattices recently observed by Weiss et al.\ \cite{Weiss2}.

Our approach implies that the quantum oscillations in antidot lattices are
closely related to
the well-known Shubnikov-de Haas oscillations. We have made this relation
explicit by
giving an analogous semiclassical treatment of Shubnikov-de Haas oscillations.
One finds that
the amplitude of Shubnikov-de Haas oscillations is of a lower power in $\hbar$
and hence larger
than that of the
quantum oscillations in chaotic systems. In this sense, the periodic-orbit
contribution
to the conductivity is a sensitive probe of quantum chaos. This is analogous to
the well-known
result that periodic-orbit contributions to the density of states are much
larger for
integrable systems (Berry-Tabor formula \cite{Berry}) than for chaotic systems
(Gutzwiller
formula \cite{Gutzwiller}).
Most importantly, the close relationship between Shubnikov-de Haas oscillations
and the
periodic-orbit contribution in chaotic systems implies that both effects should
be observable
under analogous conditions. Hence observation of the latter requires that the
elastic and
inelastic mean free paths be much larger than the length of the periodic orbit
but may be
smaller than the sample size. Furthermore, the temperature must be smaller than
$\hbar/T_\gamma$
which is the analog of the Landau-level spacing for Shubnikov-de Haas
oscillations. Both
effects also require a macroscopically homogeneous sample. Otherwise periodic
orbits in
different parts of the sample contribute with different phases resulting in a
reduced
amplitude due to destructive interference.

Our calculation of the periodic-orbit contribution to the conductivity of
antidot lattices
follows the spirit of but is different in detail from Gutzwiller's derivation
of the
periodic-orbit contributions to the density of states. Like Gutzwiller's trace
formula,
our results are strictly valid only for completely chaotic systems. Of course,
many systems
exhibit mixed dynamics with both regular and chaotic regions in phase space. In
fact, antidot
lattices tend to become mixed systems with increasing magnetic field or
increasing steepness
of the antidot potential. A quantitative theory for the periodic-orbit
contributions to the
conductivity in mixed systems remains a challenge for future work. Fortunately,
the antidot
sample which was used to demonstrate the quantum oscillations for the first
time \cite{Weiss2}
appears to be quite close to completely chaotic dynamics over a significant
magnetic-field range
so that a comparison with our theoretical results is meaningful.

It is interesting to compare our work with other recent semiclassical studies
of transport
properties. Physically, one also expects a weak-localization correction to the
classical
conductivity for the antidot lattice. It appears reasonable that the
weak-localization correction
for antidot lattices should be of the same order in Planck's constant $\hbar$
as that for
diffusive systems. This would imply that the weak-localization correction is of
the same order in
$\hbar$ as the periodic-orbit contributions computed in this paper.
Nevertheless, the
stationary-phase conditions do not allow for interference between time-reversed
trajectories.
Recently, Argaman argued in the context of diffusive systems
that a resolution of this dilemma requires one to go beyond standard
semiclassical theory. For small but finite $\hbar$ he also included
trajectories which approximately
(but not strictly) satisfy the stationary-phase conditions and was then able to
rederive the
diagrammatic results for weak localization in the diffusive regime.

Semiclassical expressions for the conductance have also been derived recently
starting from
the Landauer-B\"uttiker formula for the two-probe conductance. In particular,
semiclassical
treatments have been given for weak localization and conductance fluctuations
in phase-coherent
ballistic microstructures. In this formulation the semiclassical conductance is
given in terms of
a double sum over classical trajectories entering through one lead and exiting
through the other.
It is not clear how periodic orbits of the microstructure would enter in this
approach.
It is clearly desirable to gain a deeper understanding of the relationship
between these
different approaches to judge conclusively the range of validity of the
semiclassical results
for transport properties.

\acknowledgments{We thank D.\ Weiss for making unpublished data available to
us, and
enjoyed helpful and informative discussions with him and with H.\ Baranger, E.\
Doron, K.\
Ensslin, R.\ Gerhardts, B.\ Huckestein, R.\ Jalabert, K.\ Richter,
S.\ Tomsovic, H.\ Weidenm\"uller,
M.\ Zirnbauer, and W.\ Zwerger.}

\appendix
\section{Semiclassical Green function for electrons in a magnetic field}
\label{app:eg}
In this appendix, we derive the semiclassical Green function in Eq.\
(\ref{GreenSdH})
for an electron in the two-dimensional $x$-$y$ plane subject to a perpendicular
magnetic field.
We include a weak disorder potential resulting in the broadening
$\Gamma=\hbar/2\tau_{el}$
of the Green function defined above Eq.\ (\ref{deltagreen}). We recall that
the Green function is the Fourier transform of the propagator,
\begin{equation}
   G^+_{{\bf r}^\prime,{\bf r}}(E)={1\over i\hbar}\int_{-\infty}^\infty dt
     \exp\left\{{i\over\hbar}(E+i\Gamma)t\right\}K({{\bf r}^\prime,{\bf r}};t),
\label{appge1}
\end{equation}
where the propagator
\begin{equation}
  K({{\bf r}^\prime,{\bf r}};t)=\theta(t)\,\langle{\bf r}^\prime|\exp(-i
  {\hat H} t/ \hbar)|{\bf r}\rangle,
\label{defgreentime}
\end{equation}
is known exactly \cite{Feynman},
\begin{equation}
   K({{\bf r}^\prime,{\bf r}};t)=\theta(t)
   {m\over 2 \pi i \hbar t}{\omega_c t/2\over
   \sin(\omega_c t/2)}\exp\left\{{im\omega_c\over2\hbar}\left[{|{\bf r}-
   {\bf r}^\prime|^2\over2}\cot{\omega_c t\over2}-
   (x y^\prime-x^\prime y)\right]\right\}.
\label{greentime}
\end{equation}
To simplify notation we denote the phase of $K({\bf r}^\prime,{\bf r};t)$ by
\begin{equation}
  R({\bf r},{\bf r}^\prime;t)={m\over2}\left({|{\bf r}-
  {\bf r}^\prime|^2\over2}\cot{\omega_c t\over 2}-(xy^\prime-x^\prime
y)\right).
\label{principle}
\end{equation}
In the semiclassical limit $\hbar\to0$ the integral in Eq.\ (\ref{appge1}) can
be
performed by the stationary-phase method. The stationary-phase condition reads
\begin{equation}
  E={m\omega_c^2\over 8}\left({|{\bf r}-{\bf
r}^\prime|\over\sin(\omega_ct/2)}\right)^2.
\label{statphase}
\end{equation}
This condition has no solutions for $|{\bf r}-{\bf r}^\prime|>2R_c$ reflecting
the fact that
there are no classical trajectories between two points further apart than twice
the cyclotron
radius $R_c$. Thus the semiclassical $ G^+_{{\bf r}^\prime,{\bf r}}(E)$
vanishes for
$|{\bf r}-{\bf r}^\prime|>2R_c$. On the other hand, points with $|{\bf r}-{\bf
r}^\prime|<2R_c$
lie on two different cyclotron orbits (cf.\ Fig.\ 1), and one finds an
infinite set of stationary times $T_{n,q}$, $n=0,1,2,\ldots$,
$q=S,L$,
\begin{equation}
  T_{n,q}={2\pi n \over \omega_c}+t_q\, ,
\label{appTnq}
\end{equation}
with
\begin{eqnarray}
  t_S&=&{2\over\omega_c}\arcsin(|{\bf r}-{\bf r}^\prime|/2R_c),
  \nonumber\\
  t_L&=&{2\over\omega_c}[\pi-\arcsin(|{\bf r}-{\bf r}^\prime|/2R_c)].
\label{apptimes}
\end{eqnarray}
Here $n$ counts the number of complete cyclotron revolutions
and $q$ distinguishes between the two orbits.
The $T_{n,q}$ are the traversal times for the classical motion from ${\bf r}$
to
${\bf r}^\prime$. Performing the stationary-phase
integrals, one obtains
\begin{eqnarray}
    G^+_{{\bf r}^\prime,{\bf r}}(E)&=&{m\over2i\hbar}\left({1\over2\pi
i\hbar}\right)^{1/2}
    \sum_{n=0}^{\infty}\sum_{q=S,L}{\omega_c
    \over\sin(\omega_c T_{n,q}/2)}\,\left(\left.{\partial^2R\over\partial
    t^2}\right|_{T_{n,q}}\right)^{-1/2}
    \nonumber \\
    & & \times\exp(-T_{n,q}/2\tau_{el})\exp\left\{{i\over\hbar}
    \left[ET_{n,q}+R({\bf r},{\bf r}^\prime;T_{n,q})\right] \right\}.
\label{appgreene1}
\end{eqnarray}
One has
\begin{equation}
  \left.{\partial^2 R \over \partial t^2}\right|_{T_{n,q}}= \omega_c E \cot {
  \omega_c T_{n,q} \over 2}
\label{d2r}
\end{equation}
and replaces all trigonometric functions in the amplitude by their absolute
values. The
phases are absorbed into the Maslov indices
\widetext
\begin{eqnarray}
  \eta_{n,S} & = & 2n,
  \nonumber\\
  \eta_{n,L} & = & 2n+1.
\label{appMaslov}
\end{eqnarray}
\narrowtext
The phase appearing in Eq.\ (\ref{appgreene1}) is just the action
\begin{equation}
  S({\bf r},{\bf r}^\prime;E)=ET_{n,q}+{m\omega_c\over 2}\left[ {|{\bf r}-{\bf
r}^\prime|^2
  \over2}\cot{\omega_c T_{n,q}\over 2}-(xy^\prime-x^\prime y)\right]\, ,
\label{appaction}
\end{equation}
where the stationary times $T_{n,q}$ depend on ${\bf r}$, ${\bf r}^\prime$, and
$E$ as
described by Eqs.\ (\ref{appTnq},\ref{apptimes}). Putting everything together
one arrives
at the result given in Eq.\ (\ref{GreenSdH}).

\section{Stability amplitude}
\label{app:stability}

In this appendix a derivation of Eq.\ (\ref{monostab}) is presented. First, we
note that
evaluation of the amplitudes of the Green function (\ref{greenamp}) in the
local coordinate system
$(y,z)$ of the trajectory yields \cite{Gutzwiller}
\begin{equation}
  D_{\gamma_i}({\bf r},{\bf r}^\prime)={1\over\sqrt{|{\bf v}(z)|\,|{\bf
v}(z^\prime)|}}\left|
  {\partial^2 S_{\gamma_i}\over\partial y\partial y^\prime}\right|.
\end{equation}
Hence we need to simplify the expression
\begin{equation}
   a_\gamma=\left\{\left|{\partial^2 S_{\gamma_1}\over\partial y\partial
y^\prime}\right|
            \left.\left|{\partial^2 S_{\gamma_2}\over\partial y\partial
y^\prime}\right|
            \right/|\det F|
            \right\}^{1/2}.
\end{equation}
To do so, we introduce a matrix $A^{(i)}$ for trajectory $\gamma_i$ which
relates
to linear order small
transverse deviations in the initial and final momentum $\delta{\bf
p}^T=[\delta p,\delta
p^\prime]$ to small transverse deviations in the initial and final positions
$\delta{\bf y}^T=
[\delta y,\delta y^\prime]$,
\begin{equation}
  \delta{\bf p}=A\delta{\bf y}.
\end{equation}
In terms of the action $S_{\gamma_i}$ the matrix $A^{(i)}$ can be readily
expressed as
\begin{equation}
  A^{(i)}=\left[\begin{array}{cc}
      {-{\textstyle\partial^2 S_{\gamma_i}\over\textstyle\partial y^2}}&
      {-{\textstyle\partial^2 S_{\gamma_i}\over\textstyle\partial y\partial
y^\prime}} \\
      {\textstyle  \partial^2 S_{\gamma_i}\over\textstyle\partial
y^\prime\partial y}&
      {\textstyle\partial^2 S_{\gamma_i}\over\textstyle\partial y^{\prime2}}
\end{array}
      \right].
\end{equation}
Hence, all quantities appearing in $a_\gamma$ can be expressed in terms of
matrix
elements of $A^{(1)}$ and $A^{(2)}$. It turns out that $a_\gamma$ is more
conveniently
expressed in terms of the monodromy matrices $M$ and $N$ of the trajectories
$\gamma_1$
and $\gamma_2$, respectively. The monodromy matrix relates to linear order
the final transverse deviations
$\delta{{\bf x}^\prime}^T=[\delta y^\prime,\delta p^\prime]$ to the initial
transverse
deviations in phase space $\delta{\bf x}^T=[\delta y,\delta p]$,
\begin{equation}
  \delta{\bf x}^\prime=M\delta{\bf x}.
\end{equation}
An important property of the monodromy matrix is that it has unit determinant,
$\det M=1$.
The matrices $A^{(1)}$ and $A^{(2)}$ can now be easily written in terms of the
matrix
elements $m_{ij}$ and $n_{ij}$ of the monodromy matrices, respectively. One
finds
\begin{equation}
  A^{(1)}=\left[\begin{array}{cc}
      -{\textstyle m_{11}\over\textstyle m_{12}}&
      {\textstyle 1 \over\textstyle m_{12}} \\
      \textstyle m_{21}\!-\!{\textstyle m_{11}m_{22}\over\textstyle m_{12}}&
      {\textstyle m_{22}\over\textstyle m_{12}} \end{array}
      \right].
\end{equation}
and likewise for $A^{(2)}$. Using these relations we find for $a_\gamma$ in
terms of
the monodromy matrices $M$ and $N$,
\begin{eqnarray}
   a_\gamma^2&=&\left|{(1/m_{12})(1/n_{12})\over(m_{11}/m_{12}-n_{11}/
   n_{12})(m_{22}/m_{12}-n_{22}/ n_{12})-(1/m_{12}-1/n_{12})^2}\right|
   \nonumber\\
   &=&\left|{m_{12}n_{12}\over
n_{12}^2(m_{11}m_{22}-1)+m_{12}^2(n_{11}n_{22}-1)
   -n_{12}m_{12}(m_{11}n_{22}+n_{11}m_{22}-2)}\right|
   \nonumber\\
   &=&|n_{12}m_{21}+m_{12}n_{21}-m_{11}n_{22}-n_{11}m_{22}+2|^{-1}
   \nonumber\\
   &=&|{\rm Tr}MN^{-1}-2|^{-1}.
\end{eqnarray}
Here we used that $\det M=\det N=1$. Finally, we note that
$MN^{-1}=M_\gamma^p$, where
$M_\gamma$ is the monodromy matrix of the (primitive) periodic orbit. Hence,
one finds
the result
\begin{equation}
   a_{\gamma}=|{\rm Tr}M_\gamma^p-2|^{-1/2},
\end{equation}
which proves Eq.\ (\ref{monostab}). It is remarkable that only the stability of
the
periodic orbit enters in this factor, regardless of how long the trajectories
$\gamma_1$
and $\gamma_2$ are.

\section{Terms involving two advanced (retarded) Green functions}
\label{app:++}

In this appendix it is shown that the contributions to the conductivity
(\ref{adad})
involving two advanced or two retarded Green functions vanish in the
semiclassical
approximation. While we focus on the calculation for antidot lattices, the same
statement
holds in our treatment of Shubnikov-de Haas oscillations because the factor
displayed
in (\ref{zero}) below is also present in this case.
For definiteness, consider the contribution $\sigma_{xx}^{++}$.
Using the semiclassical expression for the Green function (\ref{Green}) one has
\begin{eqnarray}
   \sigma_{xx}^{++}= -{ie^2\over8\pi^2\hbar^2\Omega}\int d^2rd^2r^\prime
   \sum_{\gamma_1({\bf r},{\bf r}^\prime)}& &
   \sum_{\gamma_2({\bf r}^\prime,{\bf
r})}(v_1)_x(v_2^\prime)_xD_{\gamma_1}({\bf r},
   {\bf r}^\prime)D_{\gamma_2}({\bf r}^\prime,{\bf r})
   \nonumber\\
   & &\times\exp\left\{{i\over\hbar}[S_{\gamma_2}({\bf r},
   {\bf r}^\prime)+S_{\gamma_1}({\bf r}^\prime,{\bf
r})]-{i\pi\over2}[\eta_{\gamma_2}+
   \eta_{\gamma_1}]\right\}.
\end{eqnarray}
One notes that in contrast to $\sigma_{xx}^{+-}$ this contribution involves
both
trajectories running from ${\bf r}$ to ${\bf r}^\prime$ and from ${\bf
r}^\prime$ to ${\bf r}$.
In particular,
there is no classical contribution for which the phase factor cancels. Again
performing
the spatial integrals by the stationary-phase method, one finds that the
integrand
becomes stationary when ${\bf v}_1={\bf v}_2$ and ${\bf v}_1^\prime={\bf
v}_2^\prime$,
where ${\bf v}_i$ and ${\bf v}_i^\prime$ denote the velocities of $\gamma_i$ at
position
${\bf r}$ and ${\bf r}^\prime$, respectively. The stationary-phase condition is
satisfied
when both $\gamma_1$ and $\gamma_2$ are parts of the same periodic orbit. This
is illustrated
in Fig.\ 7. A straight-forward
extension of the calculation in appendix \ref{app:stability} shows that the
amplitude factors
also combine in this case to yield $|{\rm Tr}M_\gamma^p-2|^{-1/2}$ involving
only the
monodromy matrix of the periodic orbit. However, the contribution
$\sigma_{xx}^{++}$ vanishes in
the semiclassical limit because of the integrals over $z$ and $z^\prime$ along
the
periodic orbit. The analog of Eq.\ (\ref{zzprimeint}) reads
\begin{equation}
   \exp(-pT_\gamma/2\tau_{el})\sum_{n=0}^\infty
   \int_0^{T_\gamma} dt\int_0^{T_\gamma} dt^\prime\,v_x(t)v_x(t+t^\prime)\,
   \exp(-nT_\gamma/\tau_{el})
\label{zero}
\end{equation}
This expression vanishes because the exponential suppression factor due to
disorder
does not depend on $t$ or $t^\prime$ and hence the integrals are over one
period
of a periodic function with zero mean. A completely analogous argument can be
used
to show that $\sigma_{xx}^{--}$ vanishes in the semiclassical approximation.

\begin{figure}

\begin{description}

\item[Fig.\ 1] Sketch of classical trajectories which contribute to the
semiclassical
Green function $G^+_{{\bf r}^\prime,{\bf r}}(E)$ for electrons in a homogeneous
magnetic
field. The two shortest trajectories from ${\bf r}$ to ${\bf r}^\prime$ labeled
by S
(for short) and L (for long) are depicted by bold lines. There are also
contributions
from trajectories which complete additional revolutions around the cyclotron
orbits
of radius $R_c$.

\item[Fig.\ 2] Sketch of classical trajectories $\gamma_1$ and $\gamma_2$ from
${\bf r}$
to ${\bf r}^\prime$ which satisfy the stationary-phase conditions
(\ref{saddle1},
\ref{saddle2}). Both trajectories follow a periodic orbit $\gamma$ (dashed
line).
The shaded area depicts an antidot.

\item[Fig.\ 3] Experimental results of Weiss et al.\ \cite{Weiss2,Weiss3} for
quantum
oscillations $\sigma(T=0.4{\rm K})-\sigma(T=4.7{\rm K})$ as function of
magnetic field $B$
in longitudinal conductivity $\sigma_{xx}$ (full line) and Hall conductivity
$\sigma_{xy}$
(dashed line). The arrows indicate the magnetic-field values where minima in
the conductivity
are predicted according to Eq.\ (\ref{statphas}).

\item[Fig.\ 4] Numerical result for classical resistivity of the model
potential
(\ref{potential})
with $\beta=2$ as function of magnetic field $B$. The resistivity exhibits a
single
dominant peak at $B=B_0$ where the cyclotron orbit equals half the lattice
spacing of the
antidot arrays. The parameters were chosen according to the experimental
parameters
for the main sample investigated in Ref.\ \cite{Weiss2} which also showed a
single
peak in the resistivity at the same magnetic-field value. This supports the
choice
of $\beta=2$ for the steepness of the antidot potential.

\item[Fig.\ 5] (a) Poincare surface of section $(x/a,v_x/\sqrt{E_F/m})$ at
$y=0$ (mod $a$) with exponent $\beta=2$ for $R_c=a/2$ and antidot diameter
$d=a/2$
appropriate for the experimental sample in Ref.\ \cite{Weiss2}. Within our
resolution
the classical dynamics is completely chaotic.
(b) The same surface of section for $\beta=4$ and $d=a/3$. Here one finds a
mixed phase
space with both islands of regular motion and regions of chaotic motion.

\item[Fig.\ 6] Contour plot of the antidot potential (thin lines) together with
the  relevant periodic orbits for $B=B_0$ where the cyclotron radius $R_c=a/2$.
The innermost lines correspond to the boundaries of the classically-forbidden
regions.

\item[Fig.\ 7] Sketch of classical trajectories $\gamma_1$ from ${\bf
r}^\prime$
to ${\bf r}$ and $\gamma_2$ from ${\bf r}$ to ${\bf r}^\prime$ which satisfy
the
stationary-phase conditions for the terms involving two advanced or two
retarded
Green functions. Both trajectories follow
a periodic orbit $\gamma$ (dashed line). The shaded area depicts an antidot.

\end{description}

\end{figure}

\begin{table}

\begin{description}
\item[Table I] Numerical results for the traversal times $T_\gamma$, velocity
correlations
$C_\gamma(v_x,v_x)$, and stabilities $|{\rm Tr}M_\gamma-2|^{-1/2}$
of the periodic orbits shown in Fig.\ 6. The
parameters are chosen appropriately for the experimental sample in Ref.\
\cite{Weiss2}.
The magnetic field corresponds to $R_c=a/2$. For comparison, numbers are also
given
for the cyclotron orbits of unpatterned samples.
\end{description}

\end{table}

\begin{tabular}{||c|c|c|c|c||}  \hline \hline
  Orbit   &  $T_{\gamma}[10^{-12}$s] &$C_{\gamma}(v_x,v_x)$& $|{\rm Tr}
M_{\gamma}-2|^{-1/2}$\\ \hline
  $b$     &         3.42             &      1.41  &                 0.42
              \\
  $c_1$   &         4.04             &      1.62  &                 0.62
              \\
  $c_2$   &         4.07             &      1.63  &                 0.46
              \\
  $c_3$   &         4.04             &      1.62  &                 0.62
           \\
  cyclotron &       3,14             &      1.26  &                  --
           \\
  orbit    &                         &            &
           \\\hline \hline
\end{tabular}


\begin{references}
\bibitem{bdep} Note that the leading Weyl contribution to the density of states
is
independent of the magnetic field.
\bibitem{mean} The elastic mean free path for which these numbers were
calculated is
smaller than the value measured for the unpatterned sample in Ref,\
\cite{Weiss2}.
This could be due to additional scattering from the rough edges of the
antidots. This effect should be less significant for samples with smaller
antidot
diameter. Indeed, one observes that $\delta\sigma_{xx}\gg\delta\sigma_{xy}$ in
such
samples consistent with larger values of $\omega\tau_{el}$ (D.\ Weiss, private
communication). However, one should also keep in mind that this value of
$\tau_{el}$
may be an artifact of our treatment of disorder.
\end{references}
\end{document}